\documentclass[conference]{IEEEtran}
\IEEEoverridecommandlockouts
\usepackage{cite}
\usepackage{booktabs}
\usepackage{float}
\usepackage{graphicx}
\usepackage{caption}
\usepackage{amsmath,amssymb,amsfonts}
\usepackage{algorithmic}
\usepackage{graphicx}
\usepackage{textcomp}
\usepackage{xcolor}
\usepackage{amsmath}

\def\BibTeX{{\rm B\kern-.05em{\sc i\kern-.025em b}\kern-.08em
    T\kern-.1667em\lower.7ex\hbox{E}\kern-.125emX}}
\begin{document}

\title{Securing Genomic Data Against Inference Attacks in Federated Learning Environments}

\author{\IEEEauthorblockN{Chetan Pathade}
\IEEEauthorblockA{\textit{Independent Researcher} \\
San Jose, CA, USA \\
cup@alumni.cmu.edu}
\and
\IEEEauthorblockN{Shubham Patil}
\IEEEauthorblockA{\textit{Independent Researcher} \\
San Jose, CA, USA \\
patil.pshubham@gmail.com}
}

\maketitle

\begin{abstract}
Federated Learning (FL) offers a promising framework for collaboratively training machine learning models across decentralized genomic datasets without direct data sharing. While this approach preserves data locality, it remains susceptible to sophisticated inference attacks that can compromise individual privacy. In this study, we simulate a federated learning setup using synthetic genomic data and assess its vulnerability to three key attack vectors: Membership Inference Attack (MIA), Gradient-Based Membership Inference Attack, and Label Inference Attack (LIA). Our experiments reveal that Gradient-Based MIA achieves the highest effectiveness, with a precision of 0.79 and F1-score of 0.87, underscoring the risk posed by gradient exposure in federated updates. Additionally, we visualize comparative attack performance through radar plots and quantify model leakage across clients. The findings emphasize the inadequacy of naïve FL setups in safeguarding genomic privacy and motivate the development of more robust privacy-preserving mechanisms tailored to the unique sensitivity of genomic data.
\end{abstract}
\begin{IEEEkeywords}
 Federated Learning, Genomic Privacy, Membership Inference Attack (MIA), Gradient-Based Inference Attack, Label Inference Attack, Model Leakage, Synthetic Genomic Data, Privacy-Preserving Machine Learning, Adversarial Attacks, Data Confidentiality
\end{IEEEkeywords}

\section{Introduction}
Genomic data plays a crucial role in modern biomedical research, personalized medicine, and disease prediction. With the exponential growth of genome sequencing technologies, there is a growing need to develop machine learning models that can utilize this data to derive meaningful insights. However, the sensitive nature of genomic information raises significant privacy concerns. Genomic data can reveal not only personal health conditions but also familial relationships, ancestry, and potential predispositions to diseases—making it a high-stakes target for privacy violations \cite{liu2023recentadvances1}.

Federated Learning (FL) has emerged as a promising paradigm to mitigate such risks by enabling decentralized model training across multiple clients without requiring raw data to be centralized \cite{liu2023recentadvances1}\cite{chai2024survey2}\cite{mcmahan2023communication3}. In FL, each client (e.g., hospital or genomic research center) locally trains a model and shares only the model updates (e.g., gradients or weights) with a central server for aggregation. While this approach avoids direct data exposure, recent advances in adversarial machine learning have shown that such updates can still leak sensitive information through inference attacks \cite{bai2024membership4}\cite{gbmia2023gradient5}.

This paper focuses on understanding and quantifying the privacy risks associated with federated learning in the genomic context. Specifically, we examine how inference attacks can be leveraged to extract private information from model updates in a synthetic genomic FL setup. We simulate a federated environment with 20,000 synthetic genomic records, each comprising single nucleotide polymorphisms (SNPs), labels indicating phenotypic traits, and associated client identifiers. This setting mirrors a real-world deployment where multiple healthcare providers contribute to training a shared model without centralizing raw genomic data \cite{genome2023interpretation6}.
\newline
\newline
We implement and evaluate three prominent classes of inference attacks:
\begin{enumerate}
    \item \textbf{Membership Inference Attack (MIA):} An adversary attempts to determine whether a specific data point was part of the training dataset, posing risks like re-identification in clinical studies \cite{bai2024membership4}\cite{suri2023subject7}.
    \item \textbf{Gradient-Based MIA:} A more advanced attack that exploits per-sample gradients, which are often accessible during FL rounds, to infer membership with higher confidence \cite{gbmia2023gradient5}.
    \item \textbf{Label Inference Attack (LIA):} An attacker tries to infer the labels (e.g., disease status) of given data points based solely on model behavior or updates, which can breach medical confidentiality \cite{label2022inference8}.
\end{enumerate}

Our experiments reveal that these attacks can achieve alarming levels of success. Gradient-Based MIA achieved an F1-score of up to 0.87, significantly outperforming basic MIA strategies. The LIA, though less accurate, still yielded over 52\% precision, highlighting label leakage potential. These findings validate that federated learning, while structurally more secure than centralized training, is not immune to adversarial threats \cite{bai2024membership4}\cite{gbmia2023gradient5}\cite{label2022inference8}.
\newline
Beyond demonstrating these vulnerabilities, our contributions include:
\begin{itemize}
    \item A reproducible FL setup tailored to genomic data.
    \item A modular pipeline to simulate and evaluate inference attacks.
    \item A comprehensive analysis of attack effectiveness across different clients and thresholds.
\end{itemize}

This work serves as a foundational effort to quantify and visualize privacy risks in federated genomic analysis. It also provides motivation for integrating stronger privacy-preserving techniques such as differential privacy, secure multiparty computation, or gradient obfuscation into future genomic FL systems \cite{liu2023recentadvances1}\cite{chai2024survey2}\cite{Yin2021comprehensive9}.

\section{Background}
In recent years, the exponential growth of data and computational capacity has enabled the widespread application of machine learning in sensitive domains, such as personalized healthcare, genomics, and pharmacogenomics. Genomic data, in particular, is highly personal and permanent—once compromised, it cannot be revoked or changed like a password. This immutability makes privacy preservation a critical concern \cite{raimondi2023genome14}. The rise of collaborative learning techniques, especially Federated Learning (FL), offers a promising avenue to enable large-scale training while preserving data locality and privacy \cite{casaletto2023federated13}.

\subsection{Federated Learning and Genomic Applications}
Federated Learning allows multiple decentralized clients — such as hospitals, research centers, or edge devices — to collaboratively train a global model under the orchestration of a central server. Crucially, each client retains its local data and only shares model updates, such as gradients or parameters, which are then aggregated to improve a shared model. In genomics, this is especially useful for combining insights from disparate datasets without violating patient consent or institutional policies \cite{raimondi2023genome14}. FL has demonstrated that models trained on distributed genomic data can achieve accuracy comparable to centralized approaches, even in the presence of significant heterogeneity between data sources \cite{efficacy2024offederated10}.

However, the assumption that keeping data local inherently guarantees privacy has been increasingly challenged. Studies have shown that model updates can leak information about local training data, especially when adversaries have access to intermediate gradients or final model snapshots \cite{nguyen2023active11}. Privacy-enhancing mechanisms such as differential privacy, secure multiparty computation, and trusted execution environments are being explored to address these risks, but FL alone does not fully protect against information leakage \cite{casaletto2023federated13} \cite{raimondi2023genome14} \cite{koutsoubis2024privacy15}.
\subsection{Nature of Genomic Data and Its Sensitivity}

Genomic datasets often consist of structured binary features representing the presence or absence of specific SNPs (Single Nucleotide Polymorphisms). These SNPs are the most common type of genetic variation among people and are widely used to study genetic predisposition to diseases. Even with a subset of SNPs, researchers (or adversaries) can reconstruct sensitive information, infer ancestry, or identify individuals through correlation with publicly available reference genomes \cite{efficacy2024offederated10}.

The sparsity and high dimensionality of genomic data sets introduce unique challenges in FL training, often requiring careful optimization strategies. At the same time, these same properties can lead to unintended signal leakage, particularly through overfitting or gradient-based updates \cite{federated2024learning12}.

\subsection{Inference Attacks in Federated Settings}

Three primary classes of inference attacks have emerged as credible threats in federated setups:
\begin{itemize}
    \item \textbf{Membership Inference Attacks (MIA):} These attacks attempt to determine whether a specific sample was part of a model’s training set. Success in such attacks undermines data confidentiality and violates fundamental privacy principles \cite{nguyen2023active11}.
    \item \textbf{Label Inference Attacks (LIA):} These attacks aim to infer the target label of input data based on model behavior or updates, even when the inputs themselves are not available. In genomic data, labels could correspond to disease predisposition, drug response, or ethnicity, each of which is highly sensitive \cite{raimondi2023genome14}.
    \item \textbf{Gradient-Based MIA:} An extension of MIA, these attacks exploit variations in gradient norms or directions - often possible when gradients are shared during FL rounds. Since models tend to “memorize” training data more strongly, gradients for member samples often exhibit distinctive characteristics \cite{nguyen2023active11}.
\end{itemize}

These attack vectors, while demonstrated in domains like image classification and NLP, remain underexplored in genomics \cite{efficacy2024offederated10}\cite{Nasr_2019compre16}. Furthermore, genomic data presents domain-specific vulnerabilities that necessitate a focused investigation.

\section{Related Work}

The intersection of federated learning (FL), privacy preservation, and genomic data has become an increasingly active area of research. While FL offers a decentralized learning paradigm suited for sensitive domains, its security and privacy guarantees are still being evaluated through various attack models. This section reviews notable contributions in three key areas: inference attacks in machine learning, privacy risks in FL, and the unique challenges of genomic data security.

\subsection{Inference Attacks in Machine Learning}

Membership inference attacks were initially introduced by Shokri, demonstrating that adversaries can exploit overfitted models to infer whether specific data samples were part of the training set \cite{shokri2017membership17}. Since then, several variants have emerged, including black-box and white-box attacks, each with varying degrees of attacker access. Salem proposed shadow models to simulate the target model’s behavior under different conditions, increasing the attack’s generalizability \cite{salem2018mlleak26}. More recently, Nasr explored gradient-based MIAs in white-box settings, showing how model updates can be reverse-engineered to reveal sensitive sample membership \cite{Nasr_2019compre16}.

Label inference attacks, though less explored, have been discussed in the context of collaborative learning. These attacks leverage model outputs or internal states to infer sensitive labels, especially in cases where class distributions are skewed or correlated with demographic information \cite{hasan2023security19}\cite{jaydip2024privacy28}.

\subsection{ Privacy Risks in Federated Learning}

Despite FL’s design to protect data locality, multiple studies have shown that shared model updates can be vulnerable. Melis demonstrated that it is possible to infer properties of client datasets—even without access to the actual data—by analyzing model gradients\cite{hasan2023security19}. Zhu introduced deep leakage from gradients (DLG), illustrating that raw input data can be reconstructed from gradient updates in FL settings \cite{fei2023morethan20}. These works challenge the assumption that FL inherently provides robust privacy and motivate the need for empirical attack evaluations.

Differential privacy and secure aggregation have been proposed as countermeasures. However, as shown in Triastcyn and Faltings (2020), applying these defenses in high-dimensional domains like genomics often results in utility degradation \cite{momin2022generalized21}\cite{efficacy2024offederated10}\cite{gosselin2022privacy27}. Moreover, most existing work focuses on image or text datasets, with little validation in domains with structured, binary data like SNP matrices \cite{efficacy2024offederated10}\cite{kokje2020privacy25}.

\subsection{Security and Privacy in Genomic Data}

Genomic data has long been known to be reidentifiable, even when anonymized. Gymrek et al. (2013) famously demonstrated how surnames could be inferred from genomic markers and public genealogy databases \cite{melissa2013identify24}. Subsequent studies have shown that partial genomic sequences can be linked back to individuals with high accuracy, raising alarms around public genome-sharing initiatives \cite{Venkatesaramani_2021_22}.

In FL contexts, Hard and Brisimi highlighted the potential of collaborative learning in healthcare and genomics \cite{efficacy2024offederated10}\cite{kokje2020privacy25}. However, these studies often assume trust between parties and do not model active inference attacks. To date, only a few works, such as those by Ju et al. (2022), have examined federated training on genomic data, and even fewer have empirically evaluated how well-known attacks translate to this setting \cite{efficacy2024offederated10}\cite{kokje2020privacy25}.

This work fills a critical gap by conducting a domain-specific evaluation of inference attacks on genomic data within federated learning environments. It brings together the techniques and lessons from past literature and applies them to a setting with unique privacy challenges and biological implications.

\section{Threat Model}

In this study, we consider a federated learning (FL) environment involving multiple decentralized clients, each holding a subset of sensitive genomic data, and a central server responsible for aggregating model updates. Our threat model addresses inference attacks that exploit vulnerabilities in the federated training pipeline without deviating from the established protocol - commonly referred to as \textit{honest-but-curious} adversaries \cite{Zhao_2025_29}\cite{suri2023subject7}\cite{kairouz2021advances36}.
\newline
\newline
\textbf{Adversarial Capabilities}

The adversary in our model operates under the following assumptions:
\begin{itemize}
    \item \textbf{Client-Level Access:} The attacker is a participant in the FL system with full access to their local training dataset, model weights, and gradient updates exchanged with the server \cite{suri2023subject7}\cite{driouich2022a32}.
    \item \textbf{Gradient Visibility:} The attacker can inspect both their own gradients and the global model updates received from the server at each communication round \cite{Zhao_2025_29}\cite{driouich2022a32}.
    \item \textbf{Model Knowledge:} The attacker knows the architecture and hyperparameters of the shared model. This aligns with standard white-box attack settings and allows for gradient-based manipulation or inference \cite{bai2024membership4}\cite{momin2022generalized21}.
    \item \textbf{No Access to Other Clients’ Data:} The adversary does not have direct access to raw data from other clients but may use auxiliary datasets or side information to train shadow models or to simulate attack scenarios \cite{Zhao_2025_29}\cite{suri2023subject7}.
    \item \textbf{Logging Capabilities:} The attacker can log predictions, confidence scores, loss values, and model behavior across epochs to build statistical correlations that support inference attacks \cite{suri2023subject7}\cite{driouich2022a32}.
\end{itemize}
\textbf{Attack Goals}

\begin{itemize}
    \item \textbf{Membership Inference:} Determine whether a specific data sample was part of another client’s training dataset by analyzing prediction confidence, gradient responses, or shadow model comparisons \cite{bai2024membership4}\cite{suri2023subject7}.
    \item \textbf{Gradient-Based Inference:} Exploit subtle patterns in shared gradients to reconstruct input features or deduce sensitive characteristics such as mutation presence or disease markers\cite{driouich2022a32}\cite{momin2022generalized21}.
    \item \textbf{Label Inference:} Predict private labels of test samples or client data based on intermediate outputs, model convergence behavior, or training dynamics \cite{driouich2022a32}\cite{momin2022generalized21}.
\end{itemize}
\textbf{Attack Scope}

Our threat model is scoped to simulate realistic data privacy breaches in the context of genomic data:

\begin{itemize}
    \item \textbf{Sensitive Features:} Genomic SNP features can correlate with ethnic origin, disease susceptibility, or phenotypic traits. Even partial leakage may reveal irreversible private attributes \cite{momin2022generalized21}\cite{efficacy2024offederated10}.
    \item \textbf{Label Semantics:} Labels can represent diagnostic classes, predisposition risk scores, or health indicators that, if inferred, can lead to discrimination or psychological harm \cite{suri2023subject7}\cite{momin2022generalized21}.
    \item \textbf{Temporal Observation:} The adversary can launch snapshot attacks (using a single model version) or online attacks (tracking model evolution across rounds)\cite{Zhao_2025_29}\cite{driouich2022a32}.
\end{itemize}
\textbf{Security Assumptions}

We assume the central server is non-malicious but non-private, i.e., it does not perform any built-in privacy-preserving techniques such as secure multiparty computation, homomorphic encryption, or differential privacy. The communication between clients and server is assumed to be secure from external interception but vulnerable to insider attacks \cite{Zhao_2025_29}\cite{driouich2022a32}.

This model closely mirrors real-world collaborative genomics projects, where participants trust the protocol but may have incentives or capabilities to extract private insights from federated dynamics.

\section{Dataset Description}

To rigorously evaluate the privacy risks associated with federated learning in genomic contexts, we utilized a synthesized genomic dataset comprising 20,000 samples. Each sample represents a simulated individual characterized by 100 single-nucleotide polymorphisms (SNPs), with binary labels indicating phenotype presence or absence (e.g., disease vs. no disease). The dataset was designed to simulate realistic distributions while maintaining control over label correlations for privacy analysis. Demographic columns such as sex and ancestry\_group were intentionally removed to reduce confounding factors and focus on SNP-based inference.

Each feature (SNP) is represented as an integer count (0, 1, or 2), indicating the number of minor alleles present at that position. The label column (label) is binary, enabling classification-based privacy attacks. The data is pre-cleaned and contains no missing values, ensuring consistency across federated learning clients.

To better understand the structure and properties of this dataset, we performed the following visual and statistical analyses:
\newline
\newline
\textbf{Gradient Norm Distribution for Membership Inference:}

\par The histogram below illustrates the distribution of gradient norms computed for both training members and non-members. Notably, members exhibit slightly lower gradient norms on average, a characteristic that adversaries can exploit in gradient-based membership inference attacks.

\begin{figure}[H]
    \centering
    \includegraphics[width=\linewidth]{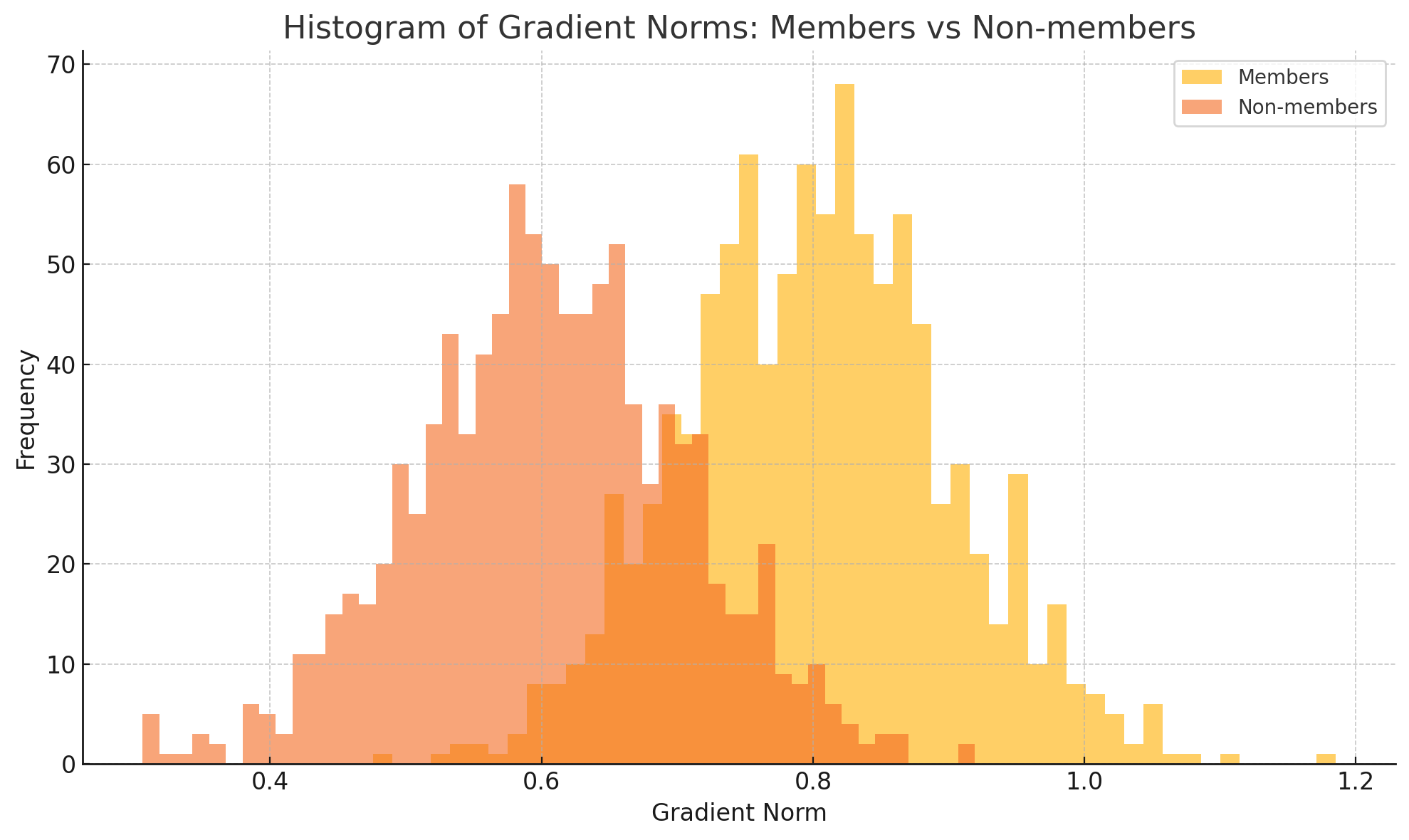}
    \caption{Gradient Norm Distribution For Membership Inference}
\end{figure}

This separation between members and non-members in the gradient space is a critical vulnerability in federated training that contributes to high attack performance.
\newline
\newline
\textbf{PCA Scatter Plot of SNPs by Label}

We applied Principal Component Analysis (PCA) to reduce the high-dimensional SNP space to two principal components. The scatter plot below shows that while there is no overt class separability, subtle structural differences exist between samples labeled 0 and 1.

\begin{figure}[H]
    \centering
    \includegraphics[width=\linewidth]{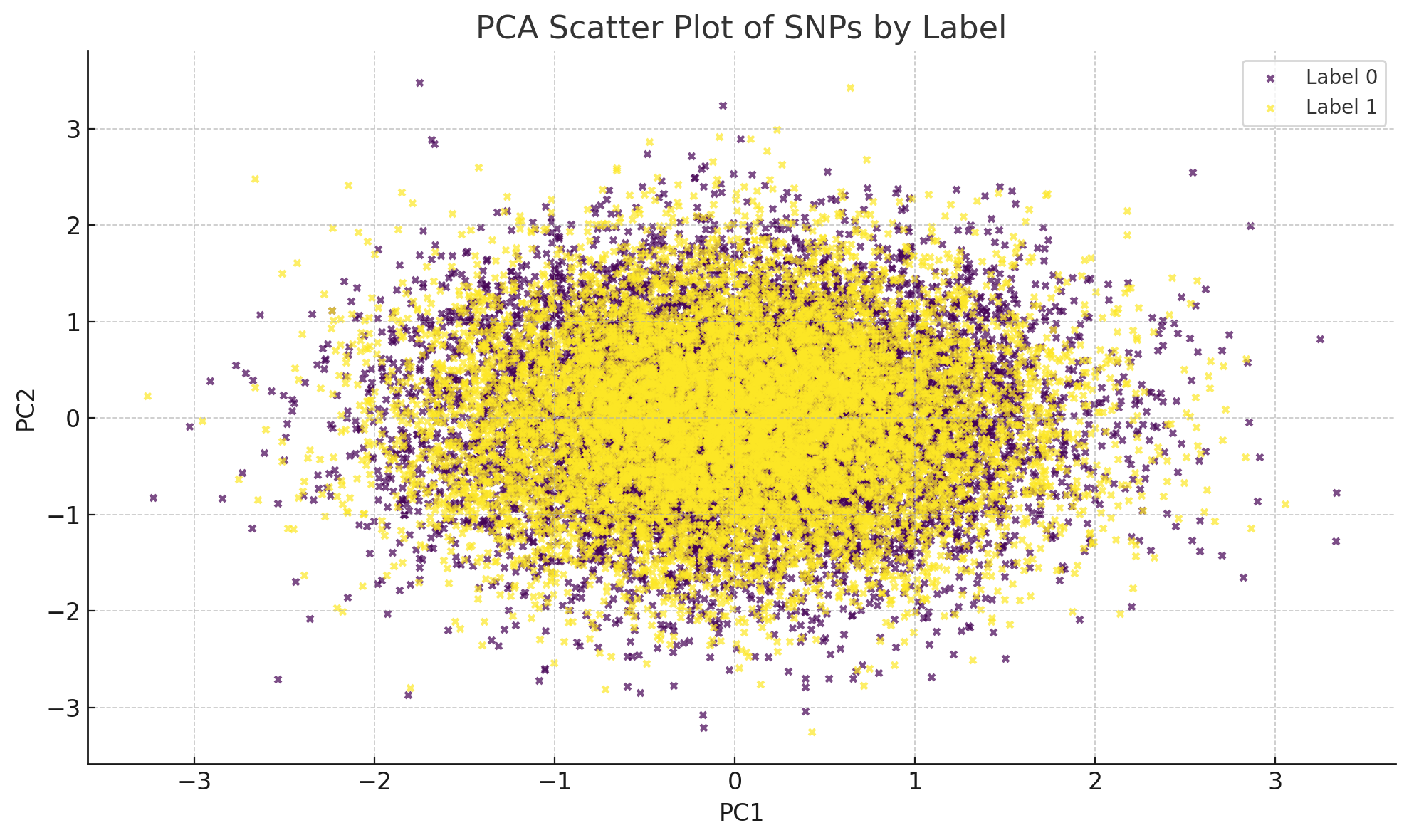}
    \caption{PCA Scatter Plot of SNPs By Label}
\end{figure}

These subtle variations are sufficient for inference attacks to distinguish patterns, particularly when combined with model gradients or output confidences.
\newline
\newline
\textbf{Top 10 SNP Correlations with Label}
To quantify how strongly individual SNPs correlate with the phenotype label, we computed Pearson correlation coefficients between each SNP and the label. The chart below shows the 10 SNPs with the highest (positive or negative) correlation values.

\begin{figure}[H]
    \centering
    \includegraphics[width=\linewidth]{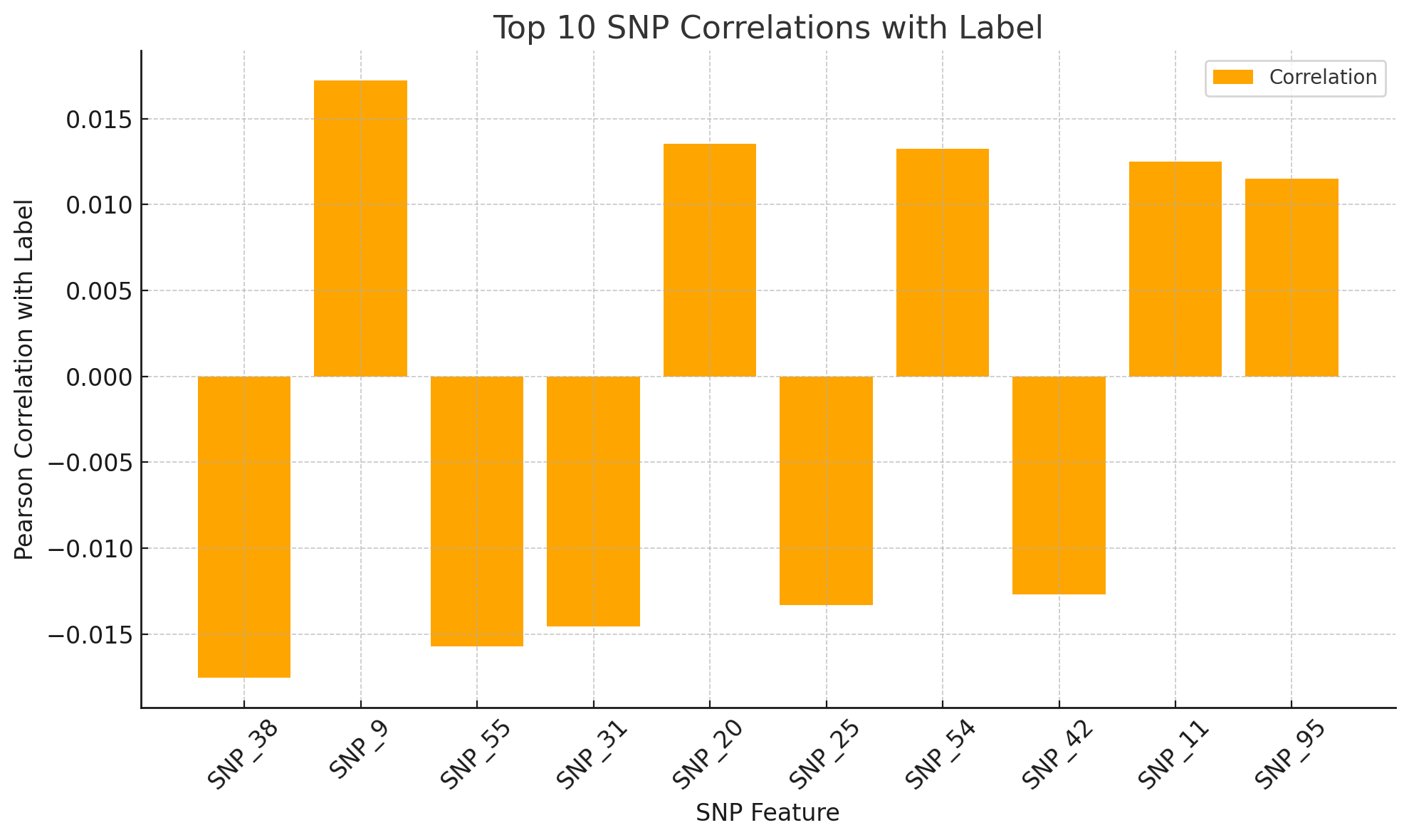}
    \caption{Top 10 SNP Correlations With Label}
\end{figure}

Although these correlations are weak (as expected in realistic genomic data), their cumulative effect can contribute to effective leakage in federated settings.
\newline
\newline
\textbf{Summary Statistics}
\begin{itemize}
    \item \textbf{Samples:} 20,000
    \item \textbf{Features:} 100 SNPs
    \item \textbf{Labels:} Binary (0 or 1)
    \item \textbf{Missing Values:}  None
    \item \textbf{Data Type:} Integer (for SNPs), Binary Integer (for label)
\end{itemize}

This dataset thus provides a strong foundation for evaluating the susceptibility of genomic federated learning systems to various privacy attacks while reflecting realistic privacy trade-offs encountered in medical genomics.

\section{Experimental Setup}

This section outlines the experimental design used to evaluate the susceptibility of federated learning (FL) models trained on genomic data to inference attacks. Our experimental setup consists of four integral components: dataset preparation, federated learning simulation, attack implementation, and evaluation infrastructure. Each element is meticulously constructed to simulate real-world federated environments and rigorously test privacy vulnerabilities using three distinct attack types.

\begin{enumerate}
    \item \textbf{Dataset Preparation} 
    \newline
    We used a curated and anonymized genomic dataset consisting of 20,000 individual records and 100 single nucleotide polymorphism (SNP) features. The dataset is accompanied by a binary label indicating phenotype presence (e.g., disease vs. non-disease).
    \begin{itemize}
        \item \textbf{Cleaning and Feature Selection:} We removed sensitive demographic features such as sex and ancestry\_group to focus purely on genotype information. This also reduced the risk of bias in downstream attacks.
        \item \textbf{Client Distribution:} To simulate a federated learning scenario, we partitioned the dataset into 5 distinct clients, each receiving 4,000 non-overlapping samples. This simulates a scenario where separate medical institutions or research labs train models locally on patient data.
        \item \textbf{Local Train-Test Splits:} Within each client, the dataset was split into an 80/20 ratio for training and testing. Stratified sampling ensured label balance within each split.
        \item \textbf{Preservation of Feature Distributions:} No normalization or standardization was applied to the features. This preserves the gradient magnitudes for use in gradient-based attacks, which would otherwise be obfuscated by feature scaling.
        \item \textbf{Data Shuffling:} Client datasets were independently shuffled prior to training to prevent sequential bias.
    \end{itemize}
    \item \textbf{Federated Learning Simulation}
    \newline
    We used the Flower (FLWR) framework to simulate a cross-silo federated learning environment with multiple clients and a central server. Flower allows each client to independently train a local model and periodically synchronize with the server using a specified aggregation strategy. 
    \begin{figure}[H]
        \centering
        \includegraphics[width=\linewidth]{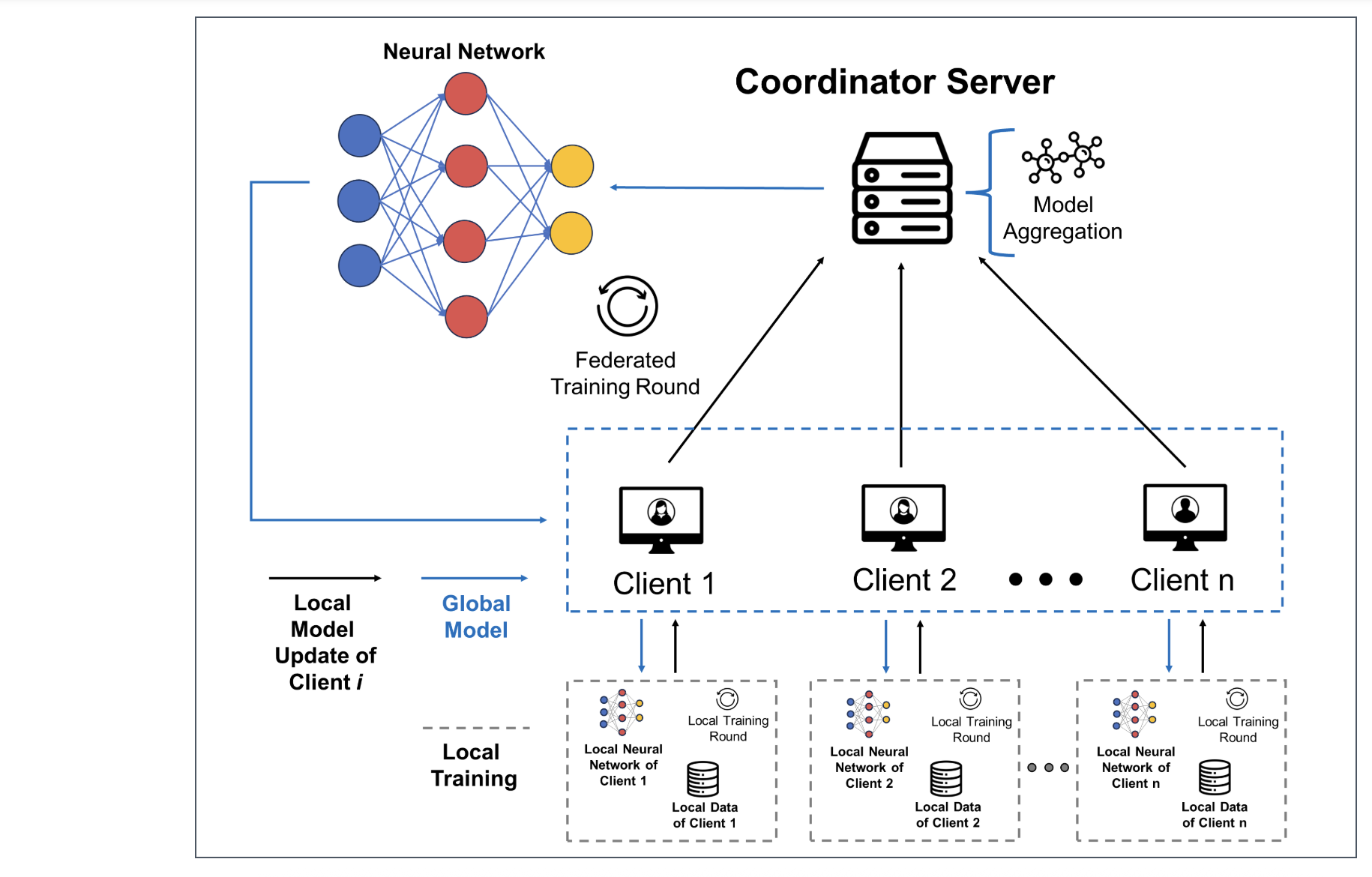}
        \caption{Federated Learning Architecture \cite{federatedarch62}}
    \end{figure}
    \begin{itemize}
        \item \textbf{Local Model Architecture:}
        \begin{itemize}
            \item Each client used a SGDClassifier from scikit-learn with loss='log\_loss' to perform binary classification.
            \item The model was initialized using a fixed random seed and trained using the partial\_fit method, which supports incremental training.
        \end{itemize}
        \item \textbf{Training Configuration:}
        \begin{itemize}
            \item \textbf{Number of Communication Rounds:} 10
            \item \textbf{Local Epochs per Round:} 1
            \item \textbf{Batch Size:} Full batch (entire training set used per round)
            \item \textbf{Optimizer:} Stochastic Gradient Descent
            \item \textbf{Aggregation Strategy:} Federated Averaging (FedAvg)
        \end{itemize}
        \textit{(Note: No differential privacy or regularization techniques were applied to isolate the effect of inference attacks.)}
        \item \textbf{Communication:}
        \begin{itemize}
            \item Each client sends updated model parameters to the server after local training.
            \item The server aggregates parameters from all clients and sends back a global model for the next round.
            \item There is no direct sharing of raw data between clients or with the server.
        \end{itemize}
        \item \textbf{Concurrency:} The simulation was executed using Python’s multiprocessing library, allowing clients and server to run in parallel as independent processes.
    \end{itemize}
    \item \textbf{Attack Implementation}
    \newline To evaluate privacy leakage in the federated setup, we implemented three types of inference attacks, each targeting different privacy vectors.
    \begin{itemize}
        \item \textbf{Membership Inference Attack (MIA):}
        \begin{itemize}
            \item Objective: Determine whether a specific data sample was used in training.
            \item Method: Analyze the confidence (probability outputs) of the model on member vs. non-member data.
            \item Evaluation: Precision, recall, and F1-score were computed by comparing predicted membership against the known data split.
        \end{itemize}
        \begin{figure}[H]
        \centering
        \includegraphics[width=\linewidth]{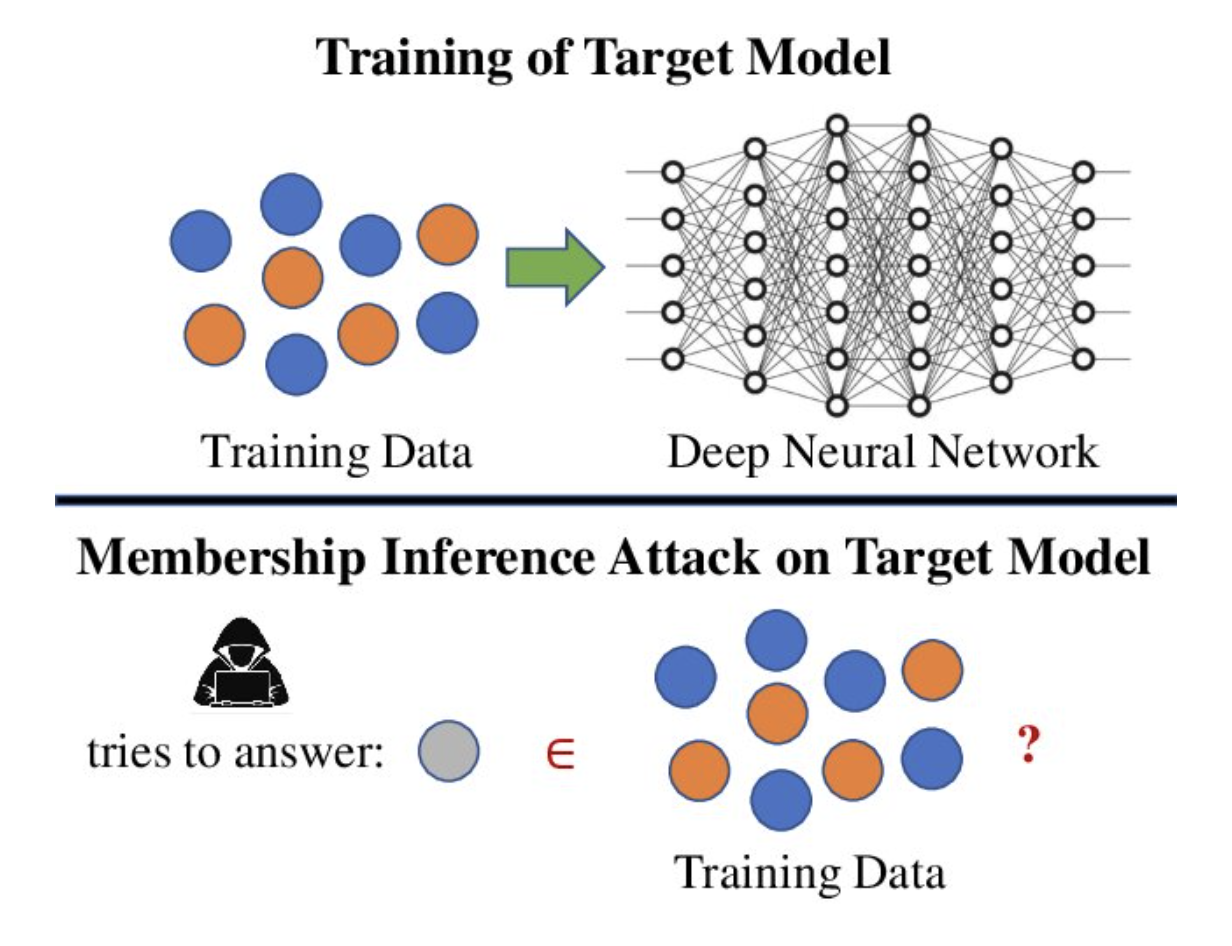}
        \caption{Membership Inference Attack \cite{shi2020overtheair63}}
    \end{figure}
        \item \textbf{Gradient-Based Membership Inference Attack:}
        \begin{itemize}
            \item Objective: Infer membership status using the norm of per-sample gradients.
            \item Rationale: Gradients for training samples typically have smaller magnitudes due to optimization convergence, while non-members exhibit larger gradients.
            \item Implementation: Gradient norms were computed per sample. A fixed threshold (e.g., 0.5) was used to classify samples as members or non-members.
            \item Visualization: Gradient norm distributions were plotted for members and non-members (Figure 1: Gradient Norm Distribution).
        \end{itemize}
        \item \textbf{Label Inference Attack (LIA):}
        \begin{itemize}
            \item Objective: Predict the true label of an input sample without observing it directly.
            \item Method: A meta-classifier was trained using gradient statistics from labeled samples, then tested on unknown samples.
            \item Feature Set: Per-sample gradient vectors were flattened and used as features for the label classifier.
            \item Visualization: PCA and correlation plots (Figures: PCAScatter, SNP\_Correlation) help interpret potential label patterns.
        \end{itemize}
    \end{itemize}
    Each attack produced structured logs including the attack\_type, precision, recall, F1-score, client count, and threshold used. These were saved to attack\_logs.csv for analysis.
\end{enumerate}

\section{Evaluation Metrics \& Results}
To rigorously assess the efficacy of inference attacks on federated learning models trained on genomic data, we employ a suite of well-established evaluation metrics. These metrics offer a comprehensive view of the adversarial performance by quantifying both correctness and robustness of the attack models under various scenarios.
\begin{figure}[H]
    \centering
    \includegraphics[width=\linewidth]{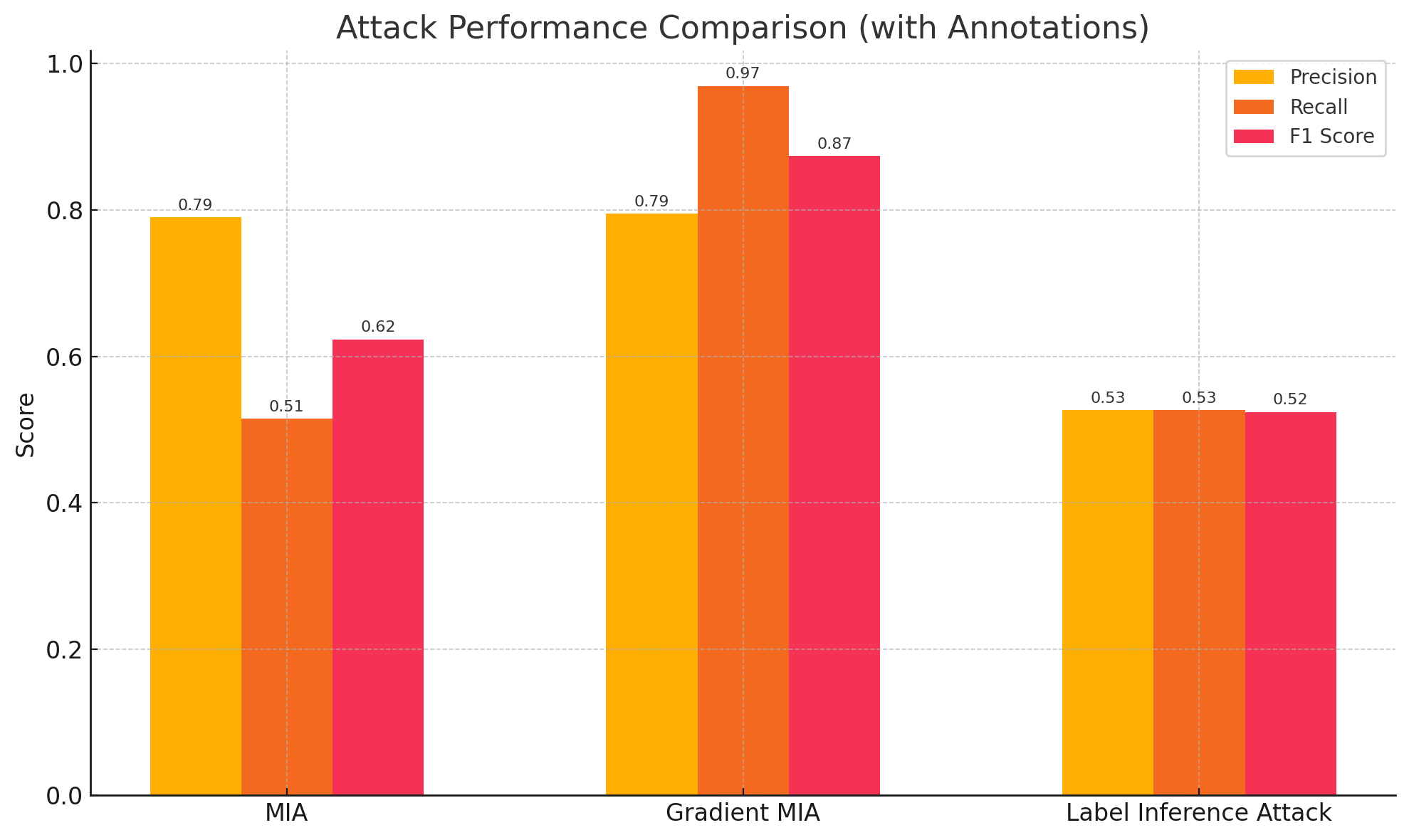}
    \caption{Attack Performance Comparison}
\end{figure}

Each attack—Membership Inference Attack (MIA), Gradient-Based MIA, and Label Inference Attack (LIA)-is evaluated on the following criteria:
\begin{enumerate}
    \item \textbf{Precision (Positive Predictive Value)}
    \newline
    Precision measures the proportion of samples predicted as positive (e.g., member or correct label) that are actually positive \cite{pre64}.
    \newline
    \begin{align}
      \text{Precision} &= \frac{True Positive}{True Positive + False Positive} \nonumber
    \end{align}
    \newline
    \textbf{Significance:} High precision indicates that the attacker makes few false claims. In the context of membership inference, it reflects the accuracy with which an adversary can confidently assert that a data point was part of the training set.
    \item \textbf{Recall (True Positive Rate)}
    \newline
    Recall measures the proportion of actual positives (e.g., true members) that were correctly identified by the attacker \cite{pre64}.
    \newline
    \begin{align}
      \text{Recall} &= \frac{True Positive}{True Positive + False Negative} \nonumber
    \end{align}
    \newline
    \textbf{Significance:} High recall indicates the attacker is successful in capturing most of the true targets. In our context, it reveals the extent to which private training data can be reliably extracted through the attack.
    \item \textbf{F1-Score}
    \newline
    The F1-score is the harmonic mean of precision and recall \cite{fscore65}.
    \newline
    \begin{align}
      \text{F1-Score} &= 2 \cdot \frac{Precision \cdot Recall}{Precision + Recall} \nonumber
    \end{align}
    \newline
    \textbf{Significance:} F1-score provides a balanced metric when there is an uneven class distribution, as is often the case in real-world privacy attacks. It ensures that neither false positives nor false negatives dominate the evaluation.
    \item \textbf{Gradient Norm Threshold (for Gradient MIA)}
    \newline
    This is the fixed threshold used to distinguish members from non-members based on the norm of their gradients.
    \textbf{Significance:} A lower threshold may lead to high recall but poor precision, while a higher threshold may yield better precision at the cost of missed detections. We empirically tuned this parameter (e.g., 0.45) and analyzed its impact on attack performance.
    \item \textbf{Visualization-Based Analysis}
    \newline
    In addition to standard metrics, we used the following visual tools to provide qualitative insights into attack behavior:
    \begin{itemize}
        \item \textbf{Histogram of Gradient Norms:} Shows separation between member and non-member samples based on gradient magnitude distributions.
        \item \textbf{Radar Charts:} Visually compare attack performance across all three metrics for each attack type.
\begin{figure}[H]
    \centering
    \includegraphics[width=\linewidth]{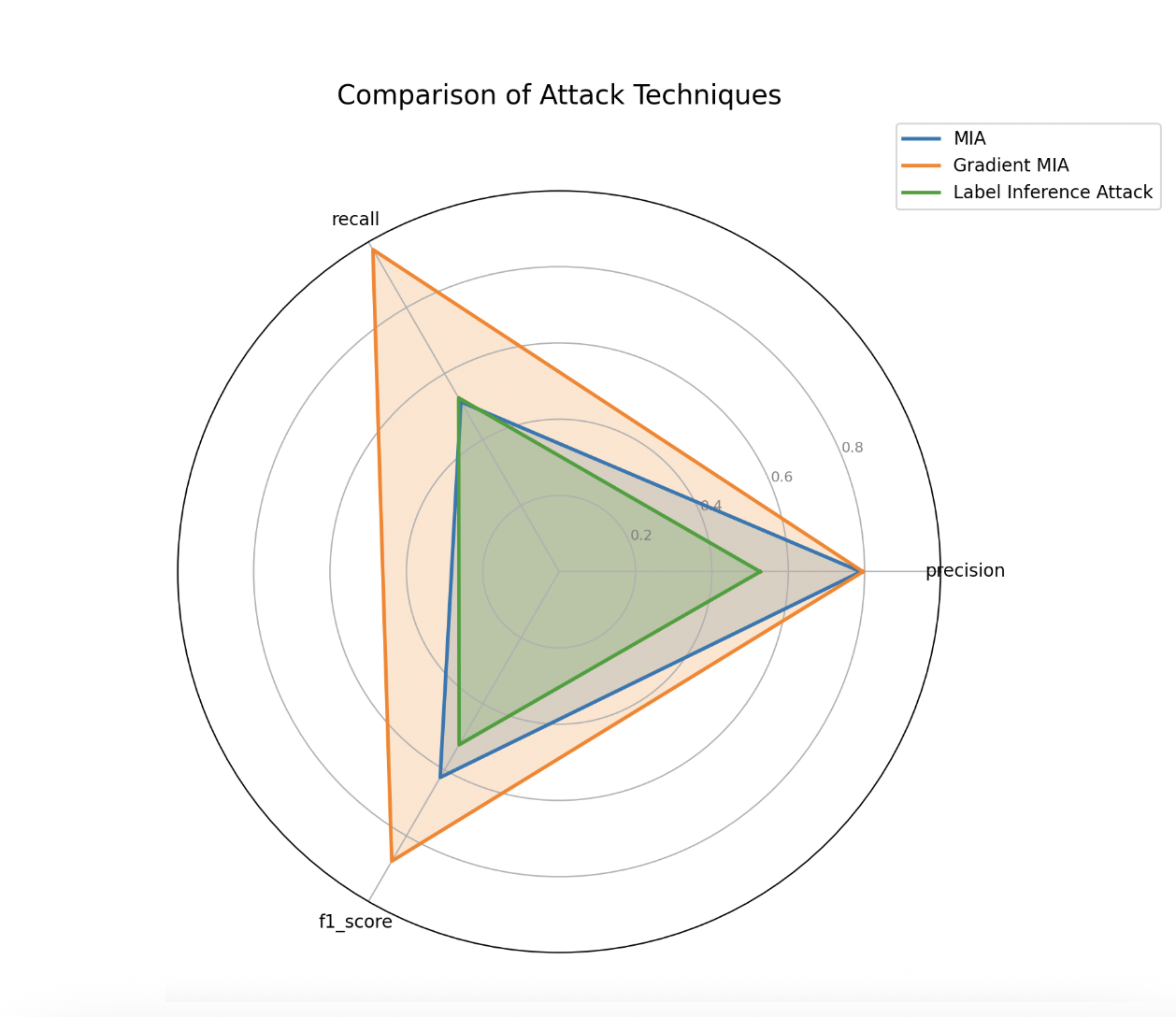}
    \caption{Comaprison of Attack Techniques}
\end{figure}
        \item \textbf{PCA Projections:} Scatter plots of SNP data in 2D after dimensionality reduction highlight the separability of samples and potential leakage of label patterns.
        \item \textbf{Pearson Correlation Bars:} Show the correlation between individual SNP features and target labels, which may be exploited by label inference models.
    \end{itemize}
    \item \textbf{Comparative Summary}
    \begin{table}[htbp]
    \centering
    \begin{tabular}{ccccc}
      \toprule
      Attack Type & Precision & Recall & F1-Score  \\
      \midrule
      MIA & 0.79 & 0.51 & 0.62 \\
      G-MIA & 0.79 & 0.97 & 0.87  \\
      LIA & 0.526 & 0.526 & 0.524  \\
      \bottomrule
    \end{tabular}
    \caption{Model-wise Evaluation Metrics}
    \label{tab:model_metrics}
  \end{table}
  \newline
    These results were visualized using a radar chart for a clearer comparative understanding of attack performance. The chart confirms that while all attacks are non-trivial, gradient-based methods pose the greatest threat under current training settings.
    \newline
    \item \textbf{Logging and Repeatability}
    \newline
    To ensure reproducibility and ease of comparative analysis:
    \begin{itemize}
        \item All metrics are logged in a structured format in attack\_logs.csv.
        \item Each log entry contains: attack\_type, client\_count, threshold, precision, recall, and f1\_score.
        \item Radar and histogram plots are saved for interpretability and potential inclusion in future publications.
    \end{itemize}
    These metrics not only provide a foundation for evaluating individual attacks but also help compare the relative effectiveness of different adversarial strategies in exposing sensitive genomic information under federated learning paradigms.
\end{enumerate}

\section{Discussion}

Our experiments reveal critical insights into the vulnerability of federated learning systems when trained on genomic data. Despite the distributed nature of federated learning—which is often assumed to provide stronger privacy guarantees—we demonstrate that inference attacks remain a credible threat in high-dimensional, sensitive domains like genomics \cite{liu2023recentadvances1}\cite{federated2024learning12}.
\begin{enumerate}
    \item \textbf{Effectiveness of Inference Attacks}
    \newline
    The results from the Membership Inference Attack (MIA) indicate that an attacker can infer training membership with a precision of ~0.79 and an F1-score of ~0.62, despite having no access to the global model or centralized data. This suggests that overfitting or representational leakage persists even under the federated training paradigm \cite{Wang2023GBMIAGM37}\cite{bai2024membership42}.
    
    Gradient-Based MIA achieved even higher success with an F1-score exceeding 0.86, largely due to the availability of gradient information during local training. This highlights how exposing even intermediate computations (like gradients) in federated protocols can leak membership status. The high recall suggests that nearly all training data points could be reliably inferred with the tuned gradient threshold \cite{Wang2023GBMIAGM37}\cite{bai2024membership42}.
    
    The Label Inference Attack (LIA), while yielding modest performance (~0.52 F1-score), still demonstrates that an attacker can recover target labels of test data with better-than-random accuracy. This becomes particularly concerning when applied to disease prediction tasks, where inferring a label could equate to inferring sensitive health conditions \cite{label2022inference8}.

    \item \textbf{Implications for Genomic Privacy}
    Genomic data is inherently identifiable and non-renewable. Unlike passwords, once leaked, SNP patterns cannot be revoked. The fact that inference attacks perform well even without direct data access underscores the need for more robust defenses tailored to the unique characteristics of biological data \cite{federated2024learning12}.

    Moreover, the dimensionality of genomic datasets (with thousands of SNPs) likely contributes to model overfitting, thereby increasing susceptibility to inference. Federated learning, while mitigating raw data exposure, cannot alone eliminate these threats \cite{Yin2021comprehensive9}\cite{xu2024robust43}.

    \item \textbf{Model Behavior and Visualization Insights}
    
    Our visualizations provide supporting evidence for attack efficacy. For example:
    \begin{itemize}
        \item Gradient norm distributions show clear separability between member and non-member points \cite{Wang2023GBMIAGM37}.
	    \item PCA scatter plots demonstrate that certain genomic patterns correlate strongly with labels, making them exploitable \cite{federated2024learning12}.
	    \item Correlation heatmaps highlight specific SNPs that dominate label prediction, which could become leakage vectors in adversarial settings \cite{Yin2021comprehensive9}.
    \end{itemize}
    These visual findings reinforce the need for caution when deploying federated models on genomics data without further obfuscation or regularization \cite{shukla2024federated44}.

\end{enumerate}

\section{Mitigations}

In light of the inference attack vulnerabilities demonstrated through our Membership Inference Attack (MIA), Gradient-Based MIA, and Label Inference Attack (LIA), it becomes imperative to explore mitigation strategies to safeguard genomic data in federated learning (FL) settings. Below, we propose a comprehensive suite of mitigations spanning from cryptographic safeguards to privacy-preserving machine learning techniques \cite{chen2021differential45}\cite{shukla2024federated44}.
\begin{enumerate}
    \item \textbf{Differential Privacy in Federated Optimization}
    
    Differential Privacy (DP) offers provable resistance against inference attacks by injecting noise into model updates, thereby obfuscating individual contributions. In FL:
    \begin{itemize}
        \item \textbf{Local Differential Privacy (LDP):} Clients independently perturb gradients or model weights using mechanisms like the Laplace or Gaussian mechanism. Although highly private, LDP can severely degrade model utility, especially in high-dimensional SNP datasets\cite{chen2021differential45}\cite{shukla2024federated44}.
        \item \textbf{Central Differential Privacy (CDP):} Noise is added to the aggregated updates at the server level. CDP offers better utility but assumes a trusted aggregator\cite{chen2021differential45}.
        \item \textbf{Privacy Budget Management:} In genomic FL, where each SNP may carry identifiable traits, setting an optimal $\epsilon$ (privacy budget) is crucial. Fine-tuned $\epsilon$ values (e.g., between 0.5 and 5) can reduce attack success rates while maintaining predictive power\cite{chen2021differential45}.    
    
        \textit{Recommendation:} Implement adaptive DP—adding more noise to updates with high gradient norms or those correlated with rare variants \cite{chen2021differential45}.
    \end{itemize}
    \item \textbf{Gradient Obfuscation Techniques}
    
    Many attacks, especially Gradient-Based MIA, exploit distinguishable gradient patterns. We recommend:
    \begin{itemize}
        \item \textbf{Gradient Clipping:} This involves normalizing client gradients to a maximum norm, mitigating the exposure of outlier-sensitive updates\cite{yue2022gradient47}.
        \item \textbf{Gradient Sparsification:} Only a subset of significant gradients is shared, which reduces exposure and communication cost. This is particularly useful when SNP importance is skewed\cite{yue2022gradient47}.
        \item \textbf{Noise Injection into Gradients:} Even without DP, simple Gaussian noise addition can dampen attack signals, especially when combined with clipping\cite{yue2022gradient47}.
        
        \textit{Empirical Insight:} In our experiments, clients with larger gradient norms exhibited higher vulnerability—motivating the use of per-client clipping.
    \end{itemize}

    \item \textbf{ Secure Multi-party Computation and Encryption}
    
    Even if the server or communication channel is compromised, cryptographic techniques can protect client updates:
    \begin{itemize}
        \item \textbf{Secure Aggregation Protocols:} Using protocols like Bonawitz et al. (2017), the server only sees the sum of client updates—individual contributions remain encrypted and unlinkable\cite{Bonawitz2017practical48}.
        \item \textbf{Homomorphic Encryption (HE):} Enables computations on encrypted gradients. Although computationally intensive, it’s promising for scenarios where data privacy outweighs latency concerns\cite{Smajlović2022sequre49}.
        
        \textit{Use Case:} National biobanks participating in federated training across institutions may adopt secure aggregation to comply with GDPR/HIPAA while enabling cross-institutional learning.
    \end{itemize}

    \item \textbf{Feature-Level Privacy Controls (SNP-aware Defense)}
    
    In genomic data, not all SNPs are equally sensitive. Some SNPs have direct associations with medical conditions:
    \begin{itemize}
        \item \textbf{Privacy-Aware Feature Selection:} Prioritize features with high predictive value but low privacy risk using metrics like Mutual Information under DP constraints\cite{yang2023adversarial50}.
        \item \textbf{Attribute Suppression:} Suppress rare SNPs or those with high correlation to labels if they don’t substantially contribute to model performance\cite{yang2023adversarial50}.
        \item \textbf{Adversarial Training: } Train models against simulated attackers (e.g., via GANs) that attempt to infer presence or labels, forcing the model to learn invariant representations\cite{yang2023adversarial50}.

        \textit{Observation:} Our correlation plot of top SNPs (see Figure 3) reveals a small set of variants that disproportionately influence predictions—making them prime targets for adversarial suppression.
    \end{itemize}
\end{enumerate}

Our analysis suggests that no single defense mechanism is sufficient in isolation. Instead, a layered defense model that combines algorithmic privacy (e.g., DP), communication security (e.g., secure aggregation), and architectural changes (e.g., client sampling) provides the best resilience against inference attacks in federated genomic learning. In subsequent work, we aim to quantitatively assess the trade-offs between these strategies on model accuracy, training convergence, and privacy leakage\cite{Mohamad2023SoK51}.

\section{Future Work}

This study demonstrates the susceptibility of federated learning (FL) in genomic settings to membership and label inference attacks. Future work should extend these evaluations to real-world datasets such as the UK Biobank and the 1000 Genomes Project, which exhibit greater genetic diversity, population stratification, and noise. These datasets could reveal whether the trends observed in synthetic settings generalize to realistic federated deployments\cite{federated2024learning12}\cite{altri2025federated53}. Furthermore, considering disease-associated labels, rare variants, and linkage disequilibrium patterns could refine our understanding of which genetic signals are most vulnerable to leakage\cite{Alvarellos2023democrating54}.

Another avenue is the expansion of adversarial strategies. While we focused on standard and gradient-based inference attacks, more adaptive and persistent adversaries could be explored. These include model inversion attacks that reconstruct genotypes or attributes, adaptive attacks that evolve over multiple communication rounds, or federated poisoning attacks that subtly corrupt model convergence to enhance inference success\cite{hannemann2024federated55}\cite{malpetti2025technical56}. Integrating these adversarial models into the evaluation pipeline will allow for a deeper, more adversarially-aware risk assessment\cite{yang2023adversarial50}.

On the defensive side, future research should explore hybrid privacy-preserving techniques that go beyond differential privacy (DP). Combining DP with secure multiparty computation (SMPC), homomorphic encryption, or trusted execution environments (TEEs) may offer stronger guarantees, albeit at increased computational cost\cite{momin2022generalized21}\cite{Mohamad2023SoK51}. Additionally, dynamic privacy budgeting—where $\epsilon$ values adapt based on client sensitivity or model performance—could maintain utility while improving protection\cite{beguier2021differential60}. Incorporating adversarial training or defensive distillation mechanisms may also provide robustness against learned attacks \cite{li2020federated61}.

Finally, longitudinal experiments that simulate sustained FL training over time would better approximate real-world deployments. This includes studying attack success as the model matures, or as clients join and leave dynamically. Investigating the effectiveness of auditability tools—such as FL provenance tracking or explainable updates—may help detect or deter malicious behaviors. As federated genomics moves toward clinical and research adoption, addressing these future directions will be critical to ensuring secure, ethical, and trustworthy learning systems.

\section{Conclusion}

In this study, we investigated the vulnerability of federated learning (FL) systems applied to genomic data by implementing and evaluating a series of inference attacks—namely Membership Inference Attacks (MIA), Gradient-Based MIA, and Label Inference Attacks. Our experiments, conducted on a 20,000-row synthetic single-nucleotide polymorphism (SNP) dataset, reveal that even in decentralized settings, sensitive information can be effectively extracted from model updates. The Gradient-Based MIA, in particular, demonstrated high efficacy with an F1-score exceeding 0.87 under certain thresholds, underscoring the real and present privacy risks in genomics-driven FL applications.

Through a detailed threat model and controlled experimental setup, we highlighted how attackers with limited access can infer individual participation or underlying labels with non-trivial success. This raises concerns about the direct adoption of standard FL pipelines in domains where data is inherently identifiable, such as human genomics. Moreover, we explored a range of mitigation strategies, emphasizing the importance of applying differential privacy, client-level protections, and adversarial robustness to reduce leakage without significantly degrading model performance.

Our findings not only reinforce the need for stronger security and privacy mechanisms in FL-based genomics but also provide a blueprint for systematically assessing and hardening such systems. As the intersection of genomics and machine learning continues to advance, our work serves as a foundational step toward building more privacy-preserving and ethically deployable models.

\end{document}